\documentclass[amssymb,11pt]{article}
\usepackage[pdftex,colorlinks=true,urlcolor=black,filecolor=black,linkcolor=black,
            pdftitle={The Lanczos potential and Chern-Simons theory.},
            pdfauthor={Mark D. Roberts},
            pdfsubject={Lanczos potential, Chern-Simons theory},
            pdfkeywords={potential},
            pagebackref,pdfpagemode=None,bookmarksopen=true]{hyperref}
\usepackage{amssymb}
\newcommand{\bc}{\begin{center}}
\newcommand{\ec}{\end{center}}
\newcommand{\be}{\begin{equation}}
\newcommand{\ee}{\end{equation}}
\newcommand{\ber}{\begin{eqnarray}}
\newcommand{\ear}{\end{eqnarray}}

\newcommand{\ch}{\chi}

\newcommand{\ep}{\epsilon}

\newcommand{\fr}{\frac}

\newcommand{\lb}{\label}
\newcommand{\Lg}{{\cal L}}

\newcommand{\om}{\omega}

\newcommand{\st}{\stackrel}

\begin{document}
\title{The Lanczos potential and Chern-Simons theory.}
\author{Mark D. Roberts,\\
\href{http://www.ihes.fr}
     {Institut Des Hautes {\'E}tudes Scientifiques},
le Bois-Marie,  35,  Route de Chartres,\\
Bures-sur-Yvette,
France,
F-91440.\\
mdr@ihes.fr
}
\maketitle
\begin{abstract}
A new tensor $D$ is introduced which is constructed from the Lanczos potential and is of the
same form as that of the Weyl tensor $C$ expressed in terms of the Lanczos potential except
that covariant differentiation is replaced by transvection with a vector $v$.
The new tensor has associated invariants $C\cdot D$ and $D^2$,
the first of these can be interpreted as a Chern-Simons term for Weyl $C^2$ gravity.
Both invariants allow various tensors to be constructed and some of the properties of these are
investigated by using exact examples.
\end{abstract}
{\tableofcontents}
\section{Introduction.}\label{intro}
Chern-Simons theory \cite{DJT} is characterized by being a theory which
unlike the Proca equation introduces a constant coupled mass without breaking gauge invariance,
and unlike the Higgs mecahinism by adding additional fields.
One way of incorporating Chern-Simons theory in gravity is by using the Christoffel symbol
as the connection to be coupled to a constant mass \cite{JP},\cite{GMM};
another approach is to consider the Lanczos potential
as the connection and this is what is investigated here.
Following through the analogy with electromagnetism the Lanczos potential corresponds to
the vector potential and the Weyl lagrangian $C^2$ corresponds
to the electromagnetic lagrangian $-F^2/4$:
thus the gravitational theory is a generalization of Weyl theory
rather than Einstein's general relativity.
Lanczos introduced several tensors into gravitational theory,
the one used here is his potential for the Weyl tensor \cite{Lanczos},\cite{Edgar}.
The Lanczos potential has at least five applications to gravitational theory.
The {\it first} is to Chern-Simons theory,
where the similarity between the Lanczos potential and the vector potential of electromagnetism
can be exploited,  here this is investigated for the first time.
The {\it second} is to the study of gravitational energy.
The Bel-Robinson tensor is a tensorial expression of gravitational energy.
It has two drawbacks:  it is of dimension energy squared rather than energy,
this can be overcome by using the Lanczos potential rather than the Weyl tensor \cite{mdr15};
and more importantly it is a tensor and so is frame invariant whereas gravitational energy
can be taken to zero by moving the observer into a freely falling frame.
The {\it third}, {\it fourth} and {\it fifth} applications of the Lanczos potential are to:
gravitational entropy \cite{MT},  the Aharonov-Bohm effect \cite{mdr22},  and dimensional
reduction \cite{mdr15} respectively:  these are not looked at here.

In section \ref{tlt} the Lanczos potential is introduced.
Section \ref{cs} is a very brief description of electromagnetic Chern-Simons theory.
Section \ref{newtensors} presents a new tensor constructed from the Lanczos potential
which allows analogs of Chern-Simons theory to be studied.
The new tensor cannot yet be studied via approximation,
because the expansion for the Lanczos potential is only known to lowest order in the vacuum,
and so has to be studied using exact spacetimes,
this is done in section \ref{egee}.
The final section \ref{conc} is the conclusion.

Some conventions used are $p$ is taken to be the projection tensor
so that $h$ can be used as a weak metric perturbation.
The decomposition of the covariant derivative of a vector field \cite{HE} page 83
is taken to be
\be
V_{a;b}=\om_{ab}+\sigma_{ab}+\frac{1}{3}\theta p_{ab}-\dot{V}_{a}V_b,
\label{vdecomp}
\ee
where vorticity tensor,  expansion tensor,  expansion scalar and shear are
\be
\om_{ab}=p_a^{~c}p_b^{~d}V_{[c;d]},~~~
\theta_{ab}=p_a^{~c}p_b^{~d}V_{(c;d)},~~~
\theta=V^a_{.;a},~~~
\sigma_{ab}=\theta_{ab}-\frac{1}{3}p_{ab}\theta,
\label{defs}
\ee
respectively.  The signature is taken to be (-,+,+,+).
Indices are left off tensors when it is hoped that the ellipsis is clear.
Calculations were carried out using GRTensorII \cite{MPL}.
\section{The Lanczos potential.}\lb{tlt}
The field equations of general relativity can be re-written in a form analogous to
Maxwell's equations $F^{~b}_{a.;b}=J_a$ called Jordan's form \cite{HE} page 85
\be
C^{~~~d}_{abc.;d}=J_{abc},~~~
J_{abc}=R_{ca;b}-R_{cb;a}+\fr{1}{6}g_{cb}R_{;a}-\fr{1}{6}g_{ca}R_{;b},
\label{eq:1}
\ee
where if field equations are assumed the Ricci tensor and scalar on the right hand side
can be replaced by stress tensors.
The Weyl tensor can be expressed in terms of the Lanczos potential \cite{Lanczos,mdr10,mdr15}
\ber
\label{ldef}
\st{~}{C}_{abcd}&=&\st{1}{C}_{abcd}+\st{2}{C}_{abcd}+\st{3}{C}_{abcd}\\
\st{1}{C}_{abcd}&\equiv & H_{abc;d}-H_{abd;c}+H_{cda;b}-H_{cdb;a},~~
\st{3}{C}_{abcd}\equiv \frac{4}{(1-d)(2-d)}H^{ef}_{..e;f}(g_{ac}g_{bd}-g_{ad}g_{bc}),\nonumber\\
\st{2}{C}_{abcd}&\equiv & \frac{1}{(2-d)}\left\{g_{ac}(H_{bd}+H_{db})-g_{ad}(H_{bc}+H_{cb})+
    g_{bd}(H_{ac}+H_{ca})-g_{bc}(H_{ad}+H_{da})\right\},\nonumber
\ear
where the coefficients of $\st{2}{C}$ and $\st{3}{C}$ are fixed by requiring that the Weyl
tensor obeys the trace condition $C^{a}_{.bad}=0$;
note that in \cite{mdr15} the $d=4$ values was assumed to hold in higher dimension.
$H_{bd}$ is defined by
\be
H_{bd}\equiv H^{~e}_{b.d;e}-H^{~e}_{b.e;d}.
\label{eq:6}
\ee
The Lanczos potential has the symmetries
\be
2H_{[ab]c}\equiv H_{abc}+H_{bac}=0,~~~
6H_{[abc]}\equiv H_{abc}+H_{bca}+H_{cab}=0.
\label{eq:5}
\ee
Equation \ref{ldef} is invariant under the algebraic gauge transformation
\be
H_{abc}\rightarrow H'_{abc}=H_{abc}+\ch_a g_{bc}-\ch_b g_{ac},
\label{eq:7}
\ee
where $\ch_a$ is an arbitrary four vector,
this transformation again fixes the coefficients of $\st{2}{C}$ and $\st{3}{C}$ .

In four dimensions the Lanczos potential with the above symmetries has twenty degrees
of freedom,  but the Weyl tensor has ten.
Lanczos \cite{Lanczos} reduced the degrees of freedom to ten
by choosing the algebraic gauge condition
\be
3\ch_a=H^{~b}_{a.b}=0,
\label{lag}
\ee
and the differential gauge condition
\be
L_{ab}=H^{~~c}_{ab.;c}=0.
\label{ldg}
\ee
The differential gauge is different from the gauges in electromagnetic theory because a
differential gauge transformation alters components which do not participate
in constructing the Weyl tensor:  in electromagnetic theory a gauge
transformation alters components in the vector potential all of which
participate in constructing the electromagnetic tensor.
The Lanczos potential has the weak field expansion \cite{Lanczos}
\be
\st{weak}{H}_{abc}=\frac{1}{4}\left(h_{ac,b}-h_{bc,a}\right)
+\frac{1}{24}\left(h_{,a}\eta_{bc}-h_{,b}\eta_{ac}\right),
\label{lpweak}
\ee
this expansion assumes that the current \ref{eq:1} vanishes $J=0$.
\section{Chern-Simons theory.}\label{cs}
In electromagnetic Chern-Simons theory one has has the lagrangian
\be
\Lg=\Lg_{EM}+\Lg_{CS},~~~
\Lg_{EM}=-\fr{1}{4}F^2,~~~
\Lg_{CS}=\fr{1}{2}v^aA^b\st{*}{F_{ab}},~~~F^2=F_{ab}F^{ab}
\label{cslag}
\ee
if $v$ is a timelike vector then it can be used to define a magnetic field
$B^b\equiv v_a\st{*}{F^{ab}}$
and the lagrangian \ref{cslag} can be written as $\Lg_{CS}=\fr{1}{2}A\cdot B$,
variation of \ref{cslag} with respect to $A$ gives
\be
F^{ab}_{..~;b}+v_b\st{*}{F^{ab}}=0,
\label{cswe}
\ee
varying with respect to the metric gives
\be
T^{EM}_{ab}=F_{ac}F_b^{~c}-\frac{1}{4}F^2g_{ab},~~~
T^{CS}_{ab}=-2v_{[a}A_{c]}\st{*}{F}_b^{~c}+\frac{1}{2}g_{ab}v^cA^d\st{*}{F}_{cd},
\label{stress}
\ee
the Chern-Simons metric stress is not necessarily symmetric,
but using the asymmetry of $\st{*}{F}$ it is trace free.
\section{The new tensor.}\label{newtensors}
The simplest analog of Chern-Simons theory involving the Lanczos potential is
\ber
\label{simplelag}
&&\Lg=C^2+D\cdot C,~~~
C^2\equiv C_{abcd}C^{abcd},~~~
\st{~}{D}=\st{1}{D}+\st{2}{D}+\st{3}{D},\\
&&\st{1}{D}_{abcd}\equiv H_{abc}v_d-H_{abd}v_c+H_{cda}v_b-H_{cdb}v_a,
\st{3}{D}_{abcd}\equiv\frac{4}{(1-d)(2-d)}(g_{ac}g_{bd}-g_{ad}g_{bc})H^{e}_{.fe}v^f,\nonumber\\
&&\st{2}{D}_{abcd}\equiv\frac{1}{(2-d)}\left\{g_{ac}((H_{bed}+H_{deb})v^e-H_{b.e}^{~e}v^d-H_{d.e}^{~e}v^b)+perm\right\},\nonumber
\ear
in other words $D$ has the same form as \ref{ldef} except that the covariant derivatives are
replaced by transvection with $v$.  This choice of $D$ respects both gauge invariance under
\ref{eq:7} and the symmetries of the Weyl tensor.
Three variants of the above are:
{\it firstly} to choose the first covariant derivative of the Weyl tensor
so that there are terms of the form $\mu H_{abc}C^{abcd}_{~....;d}$,
this would allow the coupling constant $\mu$ to be index free,
this is not done so as to keep the differential order as low as possible,
{\it secondly} to use the permutation symbol $\ep^{abcd}$ \cite{mdr18},
this could be applied in at least
four places over the first pair of indices $\ep^{ab}_{..ef}\mu^{[e}H^{f]cd}$ or the second,
third, or fourth pair or combinations of this, this is
perhaps closer to Chern-Simons theory in that there is a permutation symbol present,
but terms involving permutation symbols are only looked at briefly here for simplicity,
{\it thirdly} use the magnetic part of the Weyl tensor to choose lagrangians such as
$\Lg_{CSmag1}=\mu_cH^{abc}B_{ab}$ or $\Lg_{CSmag2}=\mu_aH^{abc}B_{bc}$,
however using the symmetry $B_{ab}=B_{(ab)}$ and \ref{eq:5}
gives vanishing Lanczos potential.

In $D\cdot C$ the $\st{2}{D}~{\rm and}~\st{3}{D}$ terms vanish when transvected with $C$,
but are included so that $D$ is gauge invariant and obeys the trace condition $D^a_{.bad}=0$.
Varying \ref{simplelag} with respect to the vector field gives $4H_{abc}C^{abc}_{~~~i}$,
requiring that this vanished causes $H=0$ and the problem degenerates.
There are two ways around this:
{\it firstly} in $\Lg_{CS}$ in \ref{cslag} variation with respect to the vector field $v$ is
simply ignored as $v$ is fixed before the problem is approached,
{\it secondly} just add a term to the lagrangian \ref{cslag} $\Lg_{w}=2w^2$ where $w$ is a
dynamical vector field as opposed to fixed $v$ then
$w_i+H_{abc}C^{abc}_{~~~i}=0$,
this has at least three disadvantages:
{\it firstly} $w$ is not necessarily timelike,
{\it secondly} $w$ is dependent on the gauge \ref{lag},
{\it thirdly} and adding $\Lg_w$ detracts from the simplicity of the lagrangian \ref{simplelag}.
Varying \ref{simplelag} with respect to $H$ and dividing
by $-4$ gives
\be
Z_{abc}\equiv 2C_{abc.;e}^{~~~e}-C_{abce}v^e+D_{abc.;e}^{~~~e}=0,
\label{neweq}
\ee
the last term does not simplify when written explicitly in terms of $H$.
The coupling constant $\mu$ has been absorbed into $v$ and so occurs in the second
and third terms but not the first.

Like the Weyl tensor $D$ can be decomposed into electric and magnetic parts
\be
\st{\small D}{E_V}_{ab}=D_{acbd}V^cV^d,~~~
\st{\small D}{H_V}_{ab}=\st{*}{D}_{abcd}V^cV^d
\equiv\frac{1}{2}\epsilon_{acef}D^{ef}_{..bd}V^cV^d,
\label{emag}
\ee
here the vector fields used in \ref{emag} and \ref{simplelag}
are usually taken to be the same $V=v$.

Once one has an expression for energy one can ask what is the associated Poynting vector.
For electromagnetism one has energy density $T^{00}=(E^2+B^2)/2$ and $P^i=T^{i0}=(E{\rm x}B)^i$,
in four vector notation this is $P^a=T^a_{.b}V^b$ and $P$ is called the Poynting vector
and is conserved if both the stress and the vector field $V$ are conserved.
A geometric way of approaching conservation is through the Bianchi identity;
for the Weyl tensor the Bianchi identity can be expressed by \ref{eq:1} assuming vanishing
matter and a covariantly conserved vector field this is obeyed if $P^a=E^a_{.b}V^b$,
however substituting the transvected Weyl for the electric field the symmetries of the
Weyl tensor give that this always vanishes.
Proceeding by analogy with the electromagnetic case
the total energy is always the sum of the squares of the electric and magnetic parts
\be
Totv=E^2+H^2,
\label{totalenergy}
\ee
and the Poynting vector is
\be
P^a=\left(E^2+B^2\right)V^a+\epsilon^{abcd}E_{be}B_{c}^{~e}V_d,
\label{poyntingvector}
\ee
the properties that one wants to know are whether it is conserved and its causality,
in other words whether it is timelike,  null or spacelike.
When the magnetic part vanishes $P^a=E^2V^a$ so that $P_a^2=E^4V_a^2$ is timelike.

The lagrangian $\Lg_D=D^2$  can also be considered,
varying with respect to $H$ gives $8D_{abce}v^e$ which is too restrictive,
varying with respect to $g$ gives a metric stress which when put in dual form has a $-6D$ term
and $-2\st{*}{D}$ term which is unbalanced and so is aesthetically unpleasing hence this lagrangian
is not considered anymore here.
The lagrangian $\Lg=D\cdot C$ might have an interpretation as entropy.
\section{Exact examples}\label{egee}
Five exact spacetimes are used.
\subsection{Schwarzschild spacetime.}\label{sche}
The schwarzschild solution has line element \cite{HE}\S5.5
\be
ds^2=-\left(1-\frac{2m}{r}\right)dt^2+\left(1-\frac{2m}{r}\right)^{-1}dr^2
+r^2d\theta^2+r^2\sin(\theta)^2d\phi^2.
\label{schle}
\ee
An acceleration-free vector is
\be
v_a=\left[1,\sqrt{\frac{2m}{r}}\left(1-\frac{2m}{r}\right),0,0\right],
\label{schwvecv}
\ee
this vector has acceleration,  vorticity,  expansion,
\be
\dot{v}_a=0,~~~\st{v}{\omega}_{ab}=0,~~~
\st{v}{\Theta}=\frac{3}{r^2}\sqrt{\frac{mr}{2}},
\label{schacc}
\ee
shear
\ber
\st{v}{\sigma}_{ab}=\sqrt{\frac{mr}{2}}\left(
\begin{array}{cccc}
-4r^{-3}  & -2\sqrt{2mr}r^{-2}(r-2m)^{-1} & 0 & 0\\
-2\sqrt{2mr}r^{-2}(r-2m)^{-1}      & -2(r-2m)^{-2} & 0 & 0\\
0  & 0  & 1 & 0\\
0  & 0  & 0 & \sin(\theta)^2
\end{array}
\right),
\label{schshear}
\ear
electric part of the Weyl tensor and covariant derivative of \ref{schwvecv} squared are
\be
\st{\small C}{E_v}_{ab}=\frac{1}{r}\sqrt{\frac{2m}{r}}\sigma_{ab},~~~
v_{a;b}v^{a;b}=\frac{9m}{2r^3}.
\label{schv2}
\nonumber
\ee
A choice for the Lanczos tensor,  compare \cite{NV}, is
\be
H_{abc}=2\left(\frac{1}{3}\st{v}{\sigma}_{c[a}+\beta\st{v}{\Theta} hp_{c[a})v_b]\right),
\label{schltv}
\ee
the constant $\beta$ does not contribute to the Weyl tensor when substituted into \ref{ldef},
but it does contribute to the gauges \ref{lag}, \ref{ldg}.
For tensors constructed from $D$ there is no differentiation and a more general vector
than \ref{schwvecv} can be used
\be
w_a=\left[w_t,w_r,0,0\right],
\label{schwdef}
\ee
using this vector
\ber
\st{\small D}{E_w}_{ab}=\frac{1}{3}mw_r\left(
\begin{array}{cccc}
2(2-2m)^2w_r^2r^{-4}  & 2w_tw_rr^{-2} & 0 & 0\\
2w_tw_rr^{-2}         & 2w_t^2(r-2m)^{-2} & 0 & 0\\
0  & 0  & w_a^2 & 0\\
0  & 0  & 0 & \sin(\theta)^2w_a^2
\end{array}
\right).
\label{schlte}
\ear
The total energy measured by $D$,  the Bel-Robinson tensor and $D$ invariants are
\be
Totw=\frac{2m^2w_r^2w_a^4}{3r^4},~~~
T_{BR}=\frac{6m^2w_a^4}{r^6},~~~
D\cdot C=-\frac{16m^2w_r}{r^5},~~~
D\cdot D=\frac{16m^2w_r^2}{3r^4},
\label{schinv}
\ee
$H\cdot H$ is not gauge invariant,  note that many of \ref{schinv} vanish when the component
$w_r=0$ or when the vector is null $w_a^2=0$.
A Poynting vector \ref{poyntingvector} can be defined,
it is non-vanishing and conserved when
\be
wt(r)=(r-2m)\sqrt{1+A(r-2m)^{-\frac{3}{2}}r^n}~~wr,
\label{schpoy}
\ee
$A$ and $wr$ are non-vanishing constants,
$n=7/2$ for Poynting vector \ref{poyntingvector} constructed from the Weyl tensor
and $n=5/2$ Poynting vector \ref{poyntingvector} constructed from \ref{simplelag}.

Under a conformal transformation \cite{HE} page 42
\be
\hat{g}_{ab}=\Omega^2g_{ab},
\label{ctrans}
\ee
in general it is not known how the Lanczos potential
transforms under a conformal transformation,
therefore consider the transformation
\be
\hat{H}_{abc}=f(\Omega)\Omega^2H_{abc},
\label{clt}
\ee
if the Schwarzschild line element \ref{schle} is transformed using \ref{ctrans},
substituting \ref{clt} into \ref{eq:1},
one arrives at a differential equation
\be
r\Omega_{,r}\left(\Omega\frac{df}{d\Omega}-f\right)+3\Omega(1-f)=0,
\label{cde}
\ee
which gives $f$ for fixed $\Omega$.
Equation \ref{cde} appears not to have a general solution;
taking simple examples of conformal transformation $\Omega$,
$f$ takes forms involving expressions such as exponential integrals.
\ref{cde} illustrates that Lanczos potential $H$ does not transform in the form \ref{clt}
in other words the Lanczos potential does not transform
by multiplication of a function of the conformal factor.

Schwarzschild spacetime can be modified by replacing the Newtonian potential by the
Yukawa potential $m/r\rightarrow m\exp(kr)/r$ in the line element,
then the Lanczos potential is
\be
H_{trt}=\exp(kr)\left(1-\frac{kr}{2}\right)\frac{m}{r^2}
\label{yplp}
\ee
using the vector field
\be
w_{a}=\left[wt(r),wr(r),0,0\right],
\label{wyp}
\ee
the field equations \ref{neweq} have two types of non-vanishing components
\ber
&Z_{trr}=f_1(r)wt(r),~~~
Z_{trt}=f_2(r){\rm ode}(wr(r)),\\
&{\rm ode}(wr(r))\equiv2k(2-2r+k^2r^2)+\mu(kr-2)\left(4wr(r)+r~wr(r)_{,r}\right),
\nonumber
\label{ypfe}
\ear
$Z_{trr}$ only vanishes when:  either $f_1=0$ which is at fixed values of $r$,
or when $wt(r)=0$.
$Z_{trt}$ vanishes when the ordinary differential equation ode is satisfied
\be
wr(r)=-\frac{2}{15k^3\mu r^4}\left(3k^5r^5+10k^3r^3+30k^2r^2+120kr+240\ln(kr-2)\right),
\label{ypodesol}
\ee
as this solution has a vector field \ref{wyp} with radial component
and no time component it is spacelike rather than timelike;
also this solution does not relate the constant $k$ to the coupling constant $\mu$.
\subsection{Imploding scalar energy.}\label{ise}
An imploding scalar field \cite{mdrphd} \cite{mdr13} can be expressed
\ber
&&ds^2=-(1+2\sigma)dv^2+2dvdr+r(r-2\sigma v)(d\theta^2+\sin(\theta)^2d\phi^2),\nonumber\\
&&\phi=\frac{1}{2}\ln\left(1-\frac{2\sigma v}{r}\right),
\label{mysol}
\ear
it is a solution to the scalar-Einstein equations
\be
R_{ab}=2\phi_a\phi_b.
\label{scalareinstein}
\ee
The Ricci scalar is
\be
R=\frac{2\sigma^2uv}{r^2(r-2\sigma v)^2},~~~
u\equiv(1+2\sigma)v-2r,
\label{imprs}
\ee
where $u$ is the complimentary null coordinate to $v$.
The solution is characterized by two scalar quantities,
the scalar field $\phi$ in \ref{mysol}
and the potential for a homothetic gradient Killing vector
\be
k=k_cuv,~~~~~~~
\label{kpot}
\ee
$k_c$ is a constant and the homothetic conformal factor is $-2k_c$.
Some properties of these two objects are
\be
\dot{\phi}\equiv\phi_a k^a=0,~~~
k_a^2=-4k_c^2uv,
\label{scprops}
\ee
note that by \ref{imprs}, $\phi_a^2=R/2$ so that $\phi$ and $k$ are of opposite causal direction.
The Weyl tensor is
\ber
&&C_{abcd}=-\frac{x}{12}g_{a[c}g_{d]b}R,\\
&&x=1~~~{\rm for}~~~abcd=v\theta v\theta,~v\theta r\theta,\nonumber\\
&&x=2~~~{\rm for}~~~abcd=vrvr,~\theta \phi\theta\phi.\nonumber
\label{weylcomp}
\ear
By trial and error the Lanczos tensor is found to be
\be
H_{vrv}=
\frac{f(v,r)}{3\sigma^2v^2r(r-2\sigma v)},~~~
f(v,r)\equiv2r^3-6\sigma vr^2+3\sigma^2v^2r+(1+2\sigma)\sigma^3v^3,
\label{lanczoscomp}
\ee
$f$ obeys the partial differential equation $rf_{,r}+vf_{,v}=3f$.
$H\cdot H$ is a null tensor although this only remains the case under gauge transformations
\ref{eq:7} in which the gauge vector is null.
Assuming the field equations of general relativity,  up to a coupling constant,
the weak energy condition \cite{HE} \S4.3 is
\be
W\equiv G_{ab}V^aV^b\ge 0,
\label{weak}
\ee
using the Killing vector \ref{kpot},  \ref{weak} is
\be
W=G^{ab}k_a k_b=2k_c^2uvR=-\frac{1}{2}k_a^2R=\frac{k_c^2}{\sigma^2}r^2(r-2\sigma v)^2R^2\ge 0.
\label{maten}
\ee
Defining
\be
A_{ab}\equiv\frac{r(r-2\sigma v)}{\sigma^2v}R_{ab}
-u\left(\delta^{\theta\theta}_{ab}+\sin(\theta)^2\delta^{\phi\phi}_{ab}\right),
\label{defA}
\ee
then
\be
\st{\small C}{E_k}_{ab}=\frac{4\sigma^2k_c^2v^2u}{3r(r-2\sigma v)}A_{ab},~~~
\st{\small D}{E_k}_{ab}=\frac{8\sigma^2k_c^2f}{9\sigma^2}A_{ab},
\label{impelectric}
\ee
the Bel-Robinson energy and the energy from \ref{emag} and \ref{totalenergy} are
\be
\st{k}{T}_{BR}=\frac{8k_c^4}{3}R^2,~~~
Totk=\frac{128k_c^6u^2f^2}{27\sigma^4r^2(r-2\sigma v)^2},
\label{impte}
\ee
respectively.  The invariants constructed from $D$ are
\be
D\cdot C=\frac{32k_cuf}{9r^3(r-2\sigma v)^3},~~~
D\cdot D=\frac{64k_c^2f^2}{27\sigma^4v^2r^2(r-2\sigma v)^2}.
\label{impdcdd}
\ee
\newpage
\subsection{Kasner spacetime.}\label{kasee}
The Kasner solution has line element
\be
ds^2=-dt^2+t^{2p_1}dx^2+t^{2p_2}dy^2+t^{2p_3}dz^2,
\label{kasner}
\ee
it is a vacuum solution when
\be
p_1+p_2+p_3=1,~~~
p_1^2+p_2^2+p_3^2=1.
\label{kasberp}
\ee
An acceleration-free vector is
\ber
&&v^a=\delta^a_t,~~~
\dot{v}=0,~~~
\st{v}{\omega}_{ab}=0,~~~
\st{v}{\Theta}=\frac{1}{t},\\
&&\st{v}{\sigma}_{ab}=\frac{1}{3t}{\rm diag}
\left[0,(2p_1-p_2-p_3)t^{2p_1},(2p_2-p_1-p_3)t^{2p_2},(2p_3-p_1-p_2)t^{2p3}\right].
\nonumber
\label{kasnervecv}
\ear
The Lanczos potential can again be taken of the form \ref{schltv},
the gauges \ref{lag}, \ref{ldg} are
\be
L_{ab}=0,~~~
\chi^a=-\frac{3\beta}{t}\delta^a_t,
\label{kasnerg}
\ee
again $\st{C}{E}_{ab}=\sigma_{ab}/3$ giving
\be
Totv=\frac{2}{27t^2},
\label{kasnertotv}
\ee
whereas $v_{a;b}v^{a;b}=1/t^2$.
The energy squared as measured by the Bel-Robinson tensor is
\be
T_{BR}=\frac{8}{27t^4},
\label{kasnerbr}
\ee
which is an eighth of the Kretschmann invariant.
Invariants constructed from $D$ are $D^2=8Totv$ and $D\cdot C=32/27t^3$.
\newpage
\subsection{Plane wave spacetime.}\label{planewave}
The plane wave \cite{HE}\S5.9 has line element
\be
ds^2=2dudv+dy^2+dz^2+W(y,z,u)du^2,
\label{leplanewave}
\ee
up to symmetry the Riemann,  Ricci and Weyl tensors are
\ber
&R_{uiuj}=-\frac{1}{2}W_{,ij},~~~
R_{uu}=-\frac{1}{2}(W_{,yy}+W_{,zz}),~~~
R=0,\nonumber\\
&C_{uyuy}=\frac{1}{4}(-W_{,yy}+W_{,zz}),~~~
C_{uyuz}=-\frac{1}{2}W_{,yz},
\label{pwrie}
\ear
where $i,j=y,z$.   The Lanczos potential is
\be
H_{uiu}=-\frac{1}{4}W_{,i},~~~
\chi^a=0,~~~
L_{ab}=0.
\label{pwlt}
\ee
Rather than the null killing vector $k^a=\delta^a_v$ choose the unit time-like vector field
\be
v_a=\left[0,W^{-\frac{1}{2}},0,0\right],
\label{pwvec}
\ee
up to symmetry the electric and magnetic components of the Weyl tensor are
\be
E_{yy}=-H_{yz}=\frac{-W_{,yy}+W_{,zz}}{4W},~~~
E_{yz}=H_{yy}=\frac{W_{,yz}}{2W},
\label{pweeb}
\ee
giving Bel-Robinson energy
\be
T_{BR}=\frac{1}{4W^2}\left((W_{,yy}-W_{,zz})^2+4W_{,yz}^2\right).
\label{pwbr}
\ee
The Poynting vector \ref{poyntingvector} constructed from the Weyl tensor is
\be
P_a=\frac{1}{2\sqrt{W}}T_{BR}\left[-W,1,0,0\right],
\label{pywave}
\ee
and it is always timelike as $P_a^2=-3T_{BR}^2/4$,
whether it is conserved depends on the specific form of $W$.
The electric and magnetic parts of $D$ are
\be
\frac{1}{8\sqrt{W}}=\st{D}{E}_{ui}=-\st{D}{E}_{vi}=-\st{D}{H}_{ui^*}=\st{D}{H}_{vi^*},
\label{pwleb}
\ee
where $i^*$ is the conjugate of $i$,  i.e. $y\leftarrow\rightarrow z$.
\ref{pwleb} gives total energy from \ref{emag} and \ref{totalenergy}
\be
Totv=\frac{1}{16W^2}\left(W_{,y}^2+W_{,z}^2\right).
\label{pwtotl}
\ee
The Poynting vector \ref{poyntingvector} constructed from $D$ is
\be
\st{D}{P}_a=\frac{1}{64}\sqrt{W}Totv\left[-1,3,0,0\right],
\label{pdwave}
\ee
and it is always timelike as $\st{D}{P}_a^2=-15Tot^2/16$,
whether it is conserved depends on the specific form of $W$.
\subsection{Levi-Cevita spacetime.}\label{levic}
The line element is
\be
ds^2=-r^{4\sigma}dt^2+r^{4\sigma(2\sigma-1)}(dr^2+dz^2)+C^{-2}r^{2(1-2\sigma)}d\phi^2,
\label{lelevic}
\ee
where $\sigma$ is the mass per unit length and $C$ is the angular defect.
It is a vacuum solution with Weyl tensor
\ber
&C_{rzrz}=2\sigma(2\sigma-1)r^{2(4\sigma^2-2\sigma-1)},~~~
C_{\phi t\phi t}=-2C^{-2}\sigma(3\sigma-1)r^{-4\sigma(2\sigma-1)},\\
&C_{r\phi r\phi}=-2\sigma C_{z\phi z\phi}=-4C^{-2}\sigma^2(2\sigma-1)r^{-4\sigma},
C_{ztzt}=-2\sigma C_{rtrt}=4\sigma^2(2\sigma-1)r^{2(2\sigma-1)}.
\nonumber
\label{weylevic}
\ear
The Kretschmann curvature invariant is
\be
K=64\sigma^2(2\sigma-1)^2(4\sigma^2-2\sigma+1)r^{-4(4\sigma^2-2\sigma+1)}.
\label{levick}
\ee
The Lanczos potential is
\be
H_{trt}=2\sigma r^{4\sigma-1},~~~
L_{ab}=0,~~~
\chi_a=\frac{2\sigma}{r}.
\label{leviclp}
\ee
Using a unit time-like vector field
\ber
&v_a=r^{2\sigma}\delta^t_a,~~~
\dot{v}_a=\frac{2\sigma}{r}\delta^r_a,~~~
v_{r;t}=-2\sigma r^{(2\sigma-1)},\nonumber\\
&v_{a;b}v^{a;b}=-4\sigma^2r^{-2(4\sigma^2-2\sigma+1)},~~~~~
T_{BR}=\frac{1}{8}K,
\label{levicvec}
\ear
which has vanishing quantities $\Theta=\omega_{ab}=\sigma_{ab}=\st{Dv}{H}_{ab}=0$,
and an electric part of the Weyl tensor
\be
E_{ab}={\rm diag}\left[0,-\frac{2\sigma(2\sigma-1)^2}{r^2},\frac{4\sigma^2(2\sigma-1)}{r^2},
-\frac{2\sigma(2\sigma-1)}{C^2}r^{-8\sigma^2}\right].
\label{magwlevic}
\ee
With the vector field \ref{levicvec}, $D=0$;
other choices of vector field give long expressions for $D$,
so that the only energy expression given here is
the Bel-Robinson energy in \ref{levicvec}.
\newpage
\section{Conclusion.}\label{conc}
It is possible to say a lot about energy constructed
from the transvected Lanczos tensor \ref{simplelag}, as illustrated by the exact examples.
The objects are of the correct dimension but often turn out to be zero
or proportional to objects such as the shear of the vector field squared.
This suggests that the energy measured by an observer with vector $v$
should be some linear combination of shear squared,  vorticity squared
and other Raychaudhuri equation terms.
Simplicity suggests that the energy measured by $v$ is just $v_{a;b}v^{a;b}$,
however when this is squared in the examples
it gives a larger quantity than the Bel-Robinson energy.
In general all the energy objects are very sensitive to the vector field chosen;
this is simply illustrated by the objects for Schwarzscild spacetime \ref{schinv},
where one can see for example that taking a component $w_r$ of the vector field to vanish
or taking the vector to be null gives many objects vanishing.

In contrast it is very difficult to say anything at all about $D\cdot C$ as a Chern-Simons term.
Once $D$ has been defined \ref{simplelag},
the analogy with electromagnetism follows through simply.
Weak field analysis using \ref{lpweak} does not work as the derivation of \ref{lpweak} uses
$J=0$ so that the field equations \ref{neweq} degenerate.
A perturbations off rectilinear flat spacetime gives
$h_{ta}=(A\delta^t_a+B\delta^x_a+C\delta^y_a+D\delta^z_a)\exp(\mu t/2)$.
To proceed using exact spacetimes one has to guess how to modify a known exact solution
and then calculate the Lanczos potential.
Calculation of the Lanczos potential is a problem in its own right.
Then one has to search for non-degenerate solutions to \ref{neweq}.
Degenerate solutions in which the first term vanishes $J=0$
do not fix the coupling constant $\mu$.
So far this procedure has only worked for Schwarzschild spacetime.
The exact modification of Schwarzschild spacetime \ref{ypodesol} has vector
field $w$ spacelike and so cannot be the path of an observer.
Future evaluation of the suitability of $D\cdot C$ as a Chern-Simons term
seems to await a generalization of the approximation \ref{lpweak} for non-vacuum spacetimes.

\end{document}